 \documentclass[preprint,floats,aps,epsfig,nofootinbib,amssymb]{revtex4}

\usepackage{slashed}
\usepackage{slashed}
\usepackage{graphicx,color}
\usepackage{epsfig}
\usepackage{subfigure}
\usepackage{epsfig}
\usepackage{dcolumn}
\usepackage{bm}
\usepackage{color}

\def\lsim{\mathrel{\rlap{\lower4pt\hbox{\hskip1pt$\sim$}}
    \raise1pt\hbox{$<$}}}         
\def\gsim{\mathrel{\rlap{\lower4pt\hbox{\hskip1pt$\sim$}}
    \raise1pt\hbox{$>$}}}         

\begin{document}

\vspace*{-5.8ex}
\hspace*{\fill}{ACFI-T15-19}

\vspace*{+3.8ex}

\title{Hiding Scalar Higgs Portal Dark Matter }

\author{Wei Chao}
\email{chao@physics.umass.edu}

 \affiliation{ Amherst Center for Fundamental Interactions, Department of Physics, University of Massachusetts-Amherst
Amherst, MA 01003, United States }

\vspace{3cm}

\begin{abstract}

We investigate a Higgs portal dark matter model by extending the Standard model (SM) with a complex singlet, $S=(s+ia )/\sqrt{2}$, where $a $ is a dark matter candidate and $S$ gets no vacuum expectation value but $s$ mixies with the SM Higgs via a trilinear interaction. 
We point out an interesting scenario, where only quartic coupling contributes to the dark matter relic abundance and there is no tree level contribution to the direct detection.  
Numerical analysis shows that the direction detection cross section, which arises at the one-loop level, is about $2\sim 5$ orders below the current LUX bound. 
Constraints from Higgs measurements as well as collider signatures of the model at the LHC are studied.

%
%
%
%

\end{abstract}

\maketitle
\section{Introduction}

The Standard Model (SM) of the particle physics is  well-established and fits perfectly with almost all the experimental observations in the elementary particle physics.
But there are several observations that the SM can not accommodate, such as neutrino masses, dark matter and the baryon asymmetry of the Universe~\cite{Agashe:2014kda}. 
Accumulating evidences point to the existence of the dark matter, which is neutral, colorless, stable and weakly interacting matter that cannot be seen with telescopes but accounts for about $25\%$ of the matter in the Universe. 
In the SM, neutrinos are stable neutral particles, which turn out too light to be the full components of the dark matter.  
Thus one needs new physics beyond the SM. 
The weakly interacting massive particle~(WIMP) is a promising dark matter candidate, since it can naturally get the observed relic density for a WIMP with mass around the electroweak scale.
Based on the interaction pattern of WIMP with the SM particle $X$, it catalyzed various $X$-portal dark matter models, of which the Higgs portal dark matter models~\cite{Patt:2006fw,Kim:2006af,MarchRussell:2008yu,Kim:2008pp,Ahlers:2008qc,Feng:2008mu,Andreas:2008xy,Barger:2008jx,Kadastik:2009ca,Piazza:2010ye,Arina:2010an,Kanemura:2010sh,Englert:2011yb,Low:2011kp,Djouadi:2011aa,Kamenik:2012hn,Gonderinger:2012rd,Lebedev:2012zw,LopezHonorez:2012kv,Okada:2012cc,Djouadi:2012zc,Bai:2012nv,Englert:2013gz,Bian:2013wna,Chang:2013lfa,Khoze:2013uia,Okada:2013bna,Fedderke:2014wda,Chao:2015uoa,Chao:2014ina,Cai:2014hka,Chao:2012pt}  became interesting and important since the discovery of the SM-like Higgs at the LHC. 
But the conventional Higgs portal dark matter is confronted with the tension between the correct dark matter relic density and constraints from underground direct detections.

To loose this tension, one needs further extensions to the minimal Higgs portal dark matter models, for example adding parity-violating effective WIMP-Higgs interactions~\cite{LopezHonorez:2012kv,Fedderke:2014wda,Chao:2015uoa}, or including extra scalar mediators~\cite{Chao:2015uoa,Chao:2014ina,Cai:2014hka,Chao:2012pt} that mixed with the SM Higgs.
In this paper, we provide insight into the Higgs portal dark matter models to study possible scenarios that have  negligible direct detection cross section. 
There are in general three types of such scenarios based on different dynamics of dark matter or Higgs.  
The first scenario is the freeze-in dark matter~\cite{Hall:2009bx}, which has a negligible initial thermal density and  feeble interactions with the thermal bath  that lead to the production of the dark matter as temperature drops below the dark matter mass.
In this scenario, the coupling of dark matter to the SM-like Higgs is so small that the direct detection cross section can be negligible.    
%
%
The second scenario is two Higgs doublets model (2HDM)-portal dark matter, in which dark matter may only interact with the scalar bilinear, $\Phi^\dagger \Phi$, where $\Phi$ is the  Higgs doublet in the Higgs basis that gets no vacuum expectation value (VEV).
In this case there is no tree-level contribution to the WIMP-nucleus scattering, thus the direct detection cross section is negligible. 
A systematic study of the phenomenology arising from this model will be given  in Ref. \cite{wchao}.
The third scenario is  hidden scalar mediated dark matter, which is the main subject of this paper. 
This scenario extends the SM with a complex scalar singlet, $S=(s+ia)/\sqrt{2}$, where $a$ is assumed to be the stable dark matter candidate, stabilized by the CP symmetry of the potential.
The muting of the direct detection cross section in this scenario originates from the speciality of the minimization condition, where  $S$ gets no VEV but $s$ mixes with the SM Higgs arising from the trilinear interaction.
As a result there might be no tree level contribution to the scattering cross section of the WIMP off nucleus and only quartic coupling is relevant to the WIMP relic density. 

We perform a systematic study to this model and find that its direct detection cross sections is about $2\sim5$ orders below the current LUX limit, providing a correct dark matter relic density.  
Although it is difficult to detect this model in underground experiments, it is still possible to search for this model at the LHC or future CEPC-SPPC collider in the mono-Higgs channel.  
Our numerical simulations show that the production cross section of $pp\to 2a+h$ at the LHC with $\sqrt{s}=13~{\rm TeV}$ is about ${\cal O} (10)~{\rm fb}$ for a ${\cal O } (100)~{\rm GeV}$ WIMP.

The remaining of this paper is organized as follows:  We briefly describe our model in section II and study  constraints from Higgs measurements in section III. Section IV is focused on the dark matter relic density and direct detection. We discuss collider signature of this model in section V. The last part is concluding remarks. We present an alternative dark matter model in the appendix.  

\section{The model}
As was discussed in the  introduction, the key point of avoiding the constraint of dark matter direct detection is introducing a hidden scalar mediator that mixes with the SM-like Higgs but gets no VEV.
In this paper, we propose two alternative scenarios (Model A and Model B), both of which may give rise to a scalar dark matter with correct relic density and escape from constraints given by underground dark matter direct detections in the meanwhile. 
The model A is more simple, we will focus on the phenomenology of it in this paper and leave the model B, which is given in the appendix, for a future study.
We extend the SM with a complex scalar singlet, $S=(s+ia)/\sqrt{2}$, which is referred as the complex scalar singlet model. $a$ is the dark matter candidate. The scalar potential can be written as
\begin{eqnarray}
V&=& -\mu_h^2 H^\dagger H + \mu_s^2 S^\dagger S+ \lambda_h (H^\dagger H)^2 + \lambda_s (S^\dagger S)^2  +\lambda_{sh}^{} (S^\dagger S) (H^\dagger H )
\nonumber \\
&& +\{  \alpha S ( H^\dagger H ) + \beta S^2 + \rho S |S|^2+ \kappa S + {\rm h.c. }   \} \; , 
\end{eqnarray}
where $H=(H^+, (h+iG+v)/\sqrt{2})^T$ is the SM Higgs.
The stability of the WIMP is guaranteed by the CP symmetry of the potential (or a partially broken $Z_2$ symmetry ), such that  $\alpha,~\beta,~\rho$ and $\kappa$ should be real.
Tadpole conditions can be written as
\begin{eqnarray}
&&\left(-\mu_h^2  + \lambda_h v_h^2 + {1\over 2 }\lambda_{sh} v_s^2   +\sqrt{2}\alpha v_s\right) v_h =0 \; , \\
&&\left\{+\mu_s^2   + \left(\lambda_s + {3 \rho \over \sqrt{2}} \right)v_s^2  +  {1 \over 2 } \lambda_{sh}  v_h^2 +2 \beta\right\}v_s  + {\alpha \over  \sqrt{2} } v_h^2 +\sqrt{2} \kappa =0  \; .
\end{eqnarray}
For the following parameter settings 
\begin{eqnarray}
+\mu_s^2   + \left(\lambda_s + {3 \rho \over \sqrt{2}} \right)v_s^2  +  {1 \over 2 } \lambda_{sh}  v_h^2 +2 \beta>0 \; , \hspace{0.5cm}\alpha v_h^2 + 2 \kappa =0 \; ,
\end{eqnarray}
one has $v_s =0$ and $v_h=246.2~{\rm GeV}$ from precision measurements. 
Even though $v_s=0$, $s$ still mixes with the SM-like Higgs $h$ from the trilinear term, with mixing angle $\theta$. The mass eigenvalues can be written as
\begin{eqnarray}
m_a^2 &=& \mu_s^2 -2\beta + {1\over 2 } \lambda_{sh} v^2 \; ,   \\ m_{s/h}^2 &=& {1\over 2} \mu_s^2 +\beta+ \lambda_h v_h^2 \pm {1\over 2 }\sqrt{ (2\lambda_h v_h^2 -\mu_s^2 -2\beta)^2 + 8 \alpha^2 v_h^2 } \; ,
\end{eqnarray}
Physical parameters of this model are $m_a, ~m_s, ~m_h, ~\theta,~ \rho,~\lambda_{sh},~\lambda_s$, where $\lambda_s, ~\rho$ and $\lambda_{sh}$ are relevant to the dark matter phenomenology,  while other parameters can be written in terms of physical parameters
\begin{eqnarray}
\lambda_h &=& {1\over 2v^2 } (m_h^2 c^2 + m_a^2 s^2) \; , \\
\mu_s^2 &=& -{1\over 2 } (m_h^2 s^2 +m_a^2 c^2 +m_a^2 ) \; , \\
\alpha &=& {cs\over \sqrt{2} v} (m_h^2 -m_a^2 ) \; , \\
\beta &=& {1\over 4 } (m_h^2 s^2 + m_a^2 c^2 -m_a^2 ) \; , \\
\kappa &=& -{1\over 2 } \alpha v_h^2 \; .
\end{eqnarray}
We are interested in the scenario where only quartic coupling contributes to the WIMP relic density, so that $\lambda_{sh}^{}$ and $\rho$ are set to be negligible  in the following study.

\section{constraints}

\begin{figure}
  \includegraphics[width=0.45\textwidth]{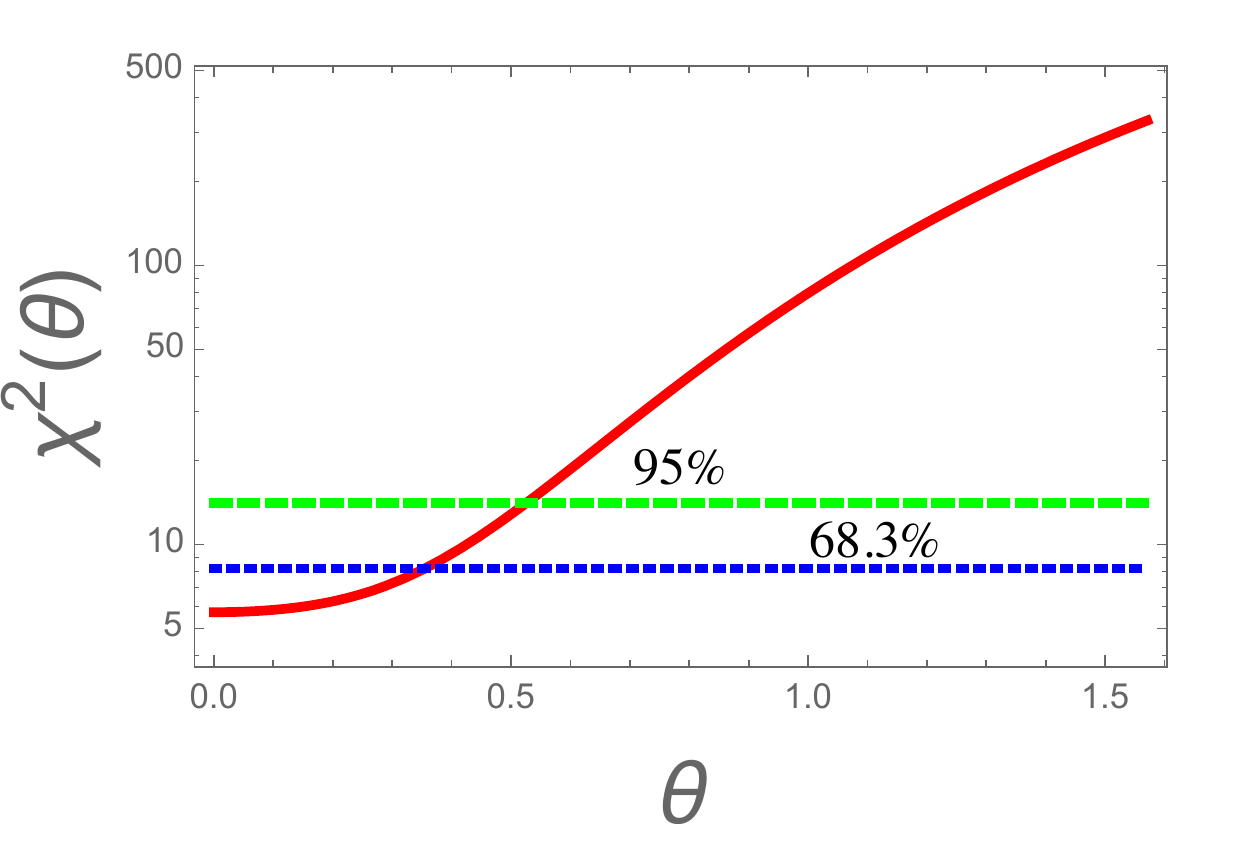}
\caption{\label{fit} The universal Higgs fit to the data from Higgs measurements at the LHC.
}
\end{figure}

%
Due to the mixing, couplings of the SM-like Higgs to all SM states are rescaled by the factor $\cos \theta$.
Such that the size of $\theta$ is constrained by the Higgs measurements at the LHC. 
The signal rates $\mu_{XX}$, which is the ratio of Higgs measurements relative to the SM-like Higgs expectations, equal to  $\cos^2\theta$.
In Ref.~\cite{Profumo:2014opa}, a global $\chi^2$ fit to the current Higgs data was performed. It gives a $95\%$ confidence level (CL) upper limit on the mixing angle: $|\theta|\leq 0.574$. 
In this paper we perform a universal Higgs fit~\cite{Giardino:2013bma} to the data given by the ATLAS and CMS collaborations. 
The fitting result is shown in Fig. \ref{fit}, where red solid line is the $\chi^2 (\theta)$,  the blue dotted and green dashed lines correspond to fixed $\chi^2 $ values at the $68\%$ and $95\%$ CL respectively. 
One has $|\theta| \leq  0.526$ at the $95\%$ CL, which is a little bit stronger than the constraint of the global $\chi^2$ fit.

\begin{figure}
  \includegraphics[width=0.45\textwidth]{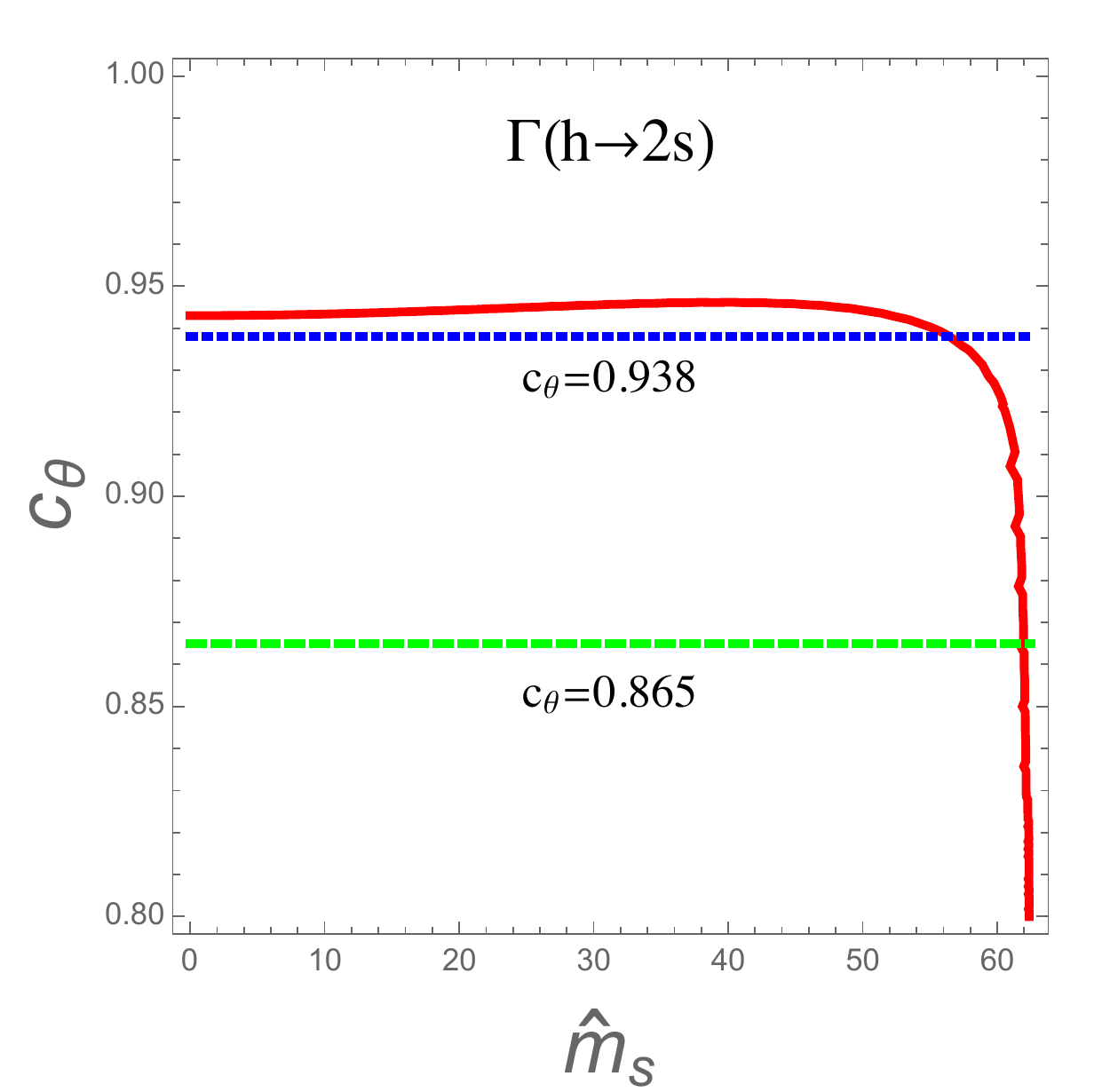}
  \includegraphics[width=0.45\textwidth]{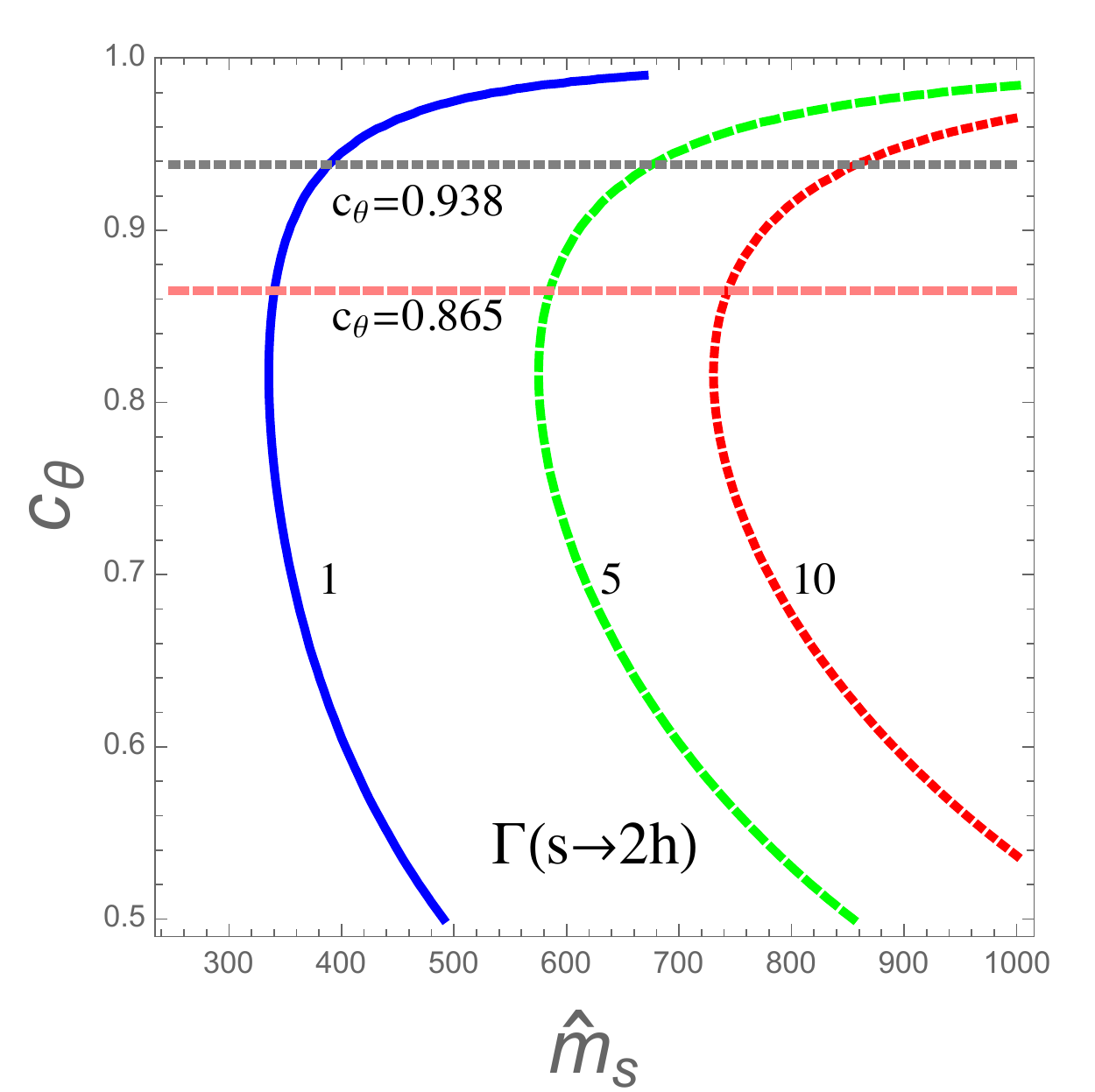}
\caption{\label{rate} Contours for the decay rates of the SM Higgs (left panel)  and the new scalar singlet (right panel) in the $m_s -c_\theta$ plane.
}
\end{figure}

For the case $ m_h > 2 m_s$, the SM Higgs can decay into $2s$.  
By assuming $\kappa_V\leq 1$, where $\kappa_V$ is the ratio of the vector boson coupling relative to the corresponding coupling in the SM, the ATLAS collaboration~\cite{Aad:2015gba} has put the upper bound on the branching ratio of $h\to 2s$: ${\rm BR} (h\to 2s)<49\%$. 
The decay rate of $h\to 2s$ can be written as
\begin{eqnarray}
\Gamma (h\to 2s )= {\sqrt{m_h^2 - 4 m_s^2 } \over 8 \pi m_h^2} \left[  (s^3 -2c^2 s) (m_h^2 -m_s^2 ){cs \over 2 v} + (m_h^2 c^2 + m_s^2 s^2 ) {3 c s^2 \over 2 v } \right ]^2  \; , 
\end{eqnarray}
where $v$ is the VEV of the SM Higgs. Notice that only two free parameters appear in the decay rate. 
We show in Fig. \ref{rate} contours of the branching ratio ${\rm BR} ( h\to 2s )$ in the $m_s-c_\theta$ plane, where the horizontal  dotted and dashed lines correspond to $c_\theta =0.938$ and $0.865$  respectively. 
The red solid line corresponds to ${\rm BR} (h\to 2 s ) = 0.49$. 
It is clear that Higgs to invisible decay mode puts more severe constraint on the mixing angle, which,  translating into the upper bound to $c_\theta$,  has $c_\theta >0.945$ for $m_s<45~{\rm GeV}$.

For the case $m_s> 2 m_h$, $s$ can decay into $2h$. The relevant decay rate is
\begin{eqnarray}
\Gamma (s\to 2h) = {\sqrt{m_s^2 -4 m_h^2} \over 8 \pi m_s^2 } \left[  (c^3-2cs^2) {cs\over 2 v } (m_s^2 -m_h^2 )+{3c^2 s \over 2 v } (m_h^2 c^2 +m_s^2 s^2)\right]^2 \; .
\end{eqnarray}  
We plot in the right panel of Fig. \ref{rate} contours of the rate $\Gamma(s\to 2h)$ in the $m_s - c_\theta $ plane, where the horizontal  dotted and dashed lines correspond to constraints on the $c_\theta$ from universal Higgs fit at the $68\%$ and $95\%$ CL respectively. 
The solid, dotted and dashed lines correspond to $\Gamma(s\to 2h)=1~{\rm GeV},~5~{\rm GeV},~10~{\rm GeV}$ respectively.
This decay rate plays very important rule in the search of heavy Higgs at the LHC through the diHiggs channel~\cite{No:2013wsa}. 
As can be seen from the right panel of Fig. 2, $\Gamma(s\to 2h)$ can not be arbitrarily large in this model. 
We refer the reader to Ref.~\cite{Arkani-Hamed:2015vfh} for the search of heavy Higgs at the future 100~TeV pp collider.

\section{dark matter}
%
%
%

%

\begin{figure}
{
   \includegraphics[width=0.25\textwidth]{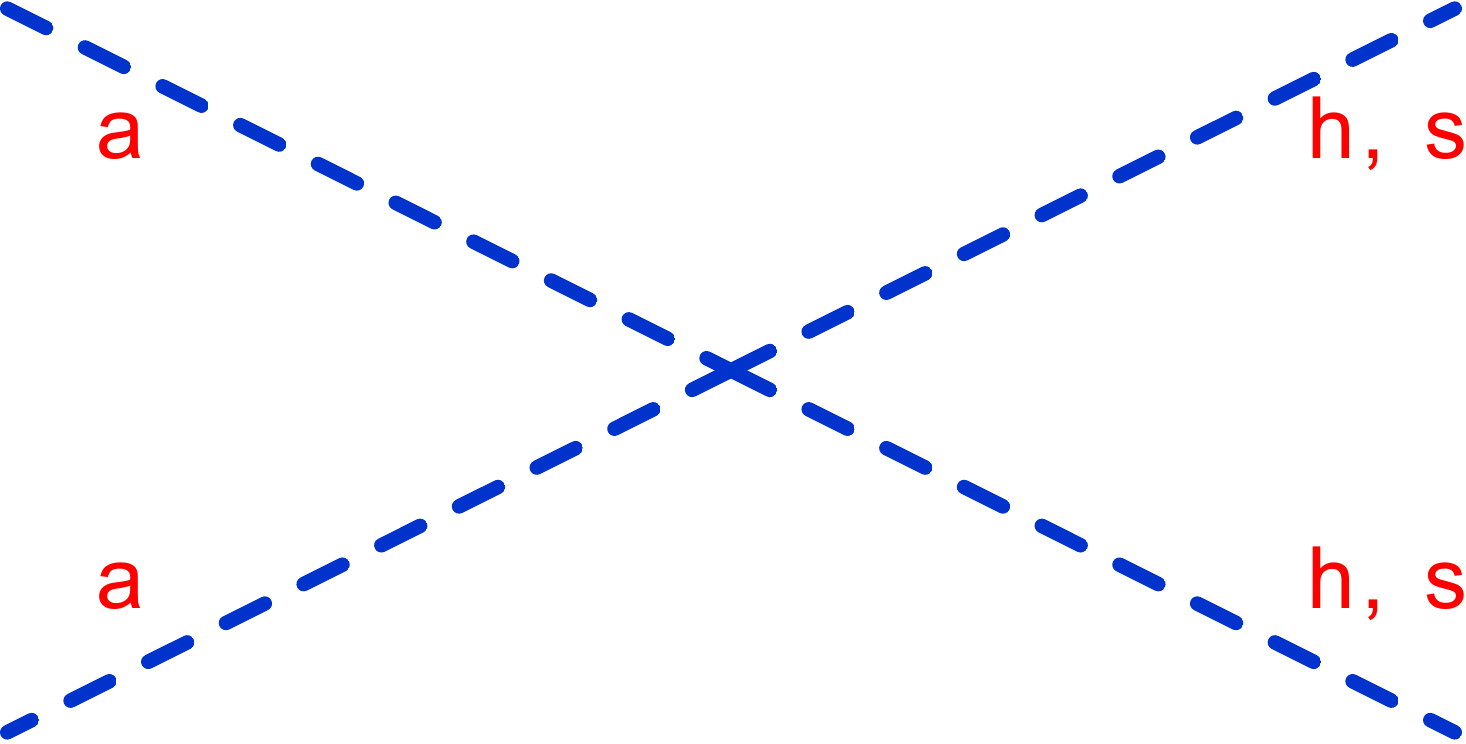}
}
\caption{\label{ffplots} Feynman diagrams relevant to the annihilation.
}
\end{figure}
%
%
The pseudo-scalar $a$ is cold dark matter candidate, which is in the thermal equilibrium in the early universe and freezes out as the temperature drops down.    
By setting $\lambda_{sh} =\rho=0$, the dark matter only interact with $s$ and $h$ through the vertex given in Fig. \ref{ffplots}. 
The evolution of the dark matter number density $n$, is governed by the Boltzmann equation:
\begin{eqnarray}
\dot{n}+3H n = -\langle \sigma v_{\rm M\slashed{o}ller} \rangle (n^2 -n_{\rm EQ}^2)
\end{eqnarray}
where $\sigma$ is the total annihilation cross section, $v_{\rm M\slashed{o}ller} =(|v_1-v_2|^2-|v_1\times v_2|^2)^{1/2}$ being the M$\slashed{o}$ller velocity, brackets denote thermal average.
It has been shown that $\langle \sigma v_{\rm M\slashed{o}ller} \rangle =\langle \sigma v_{\rm lab} \rangle = 1/2 [1 + K_1^2 (x) /K_2^2 (x)] \langle \sigma v_{\rm cm} \rangle$~\cite{Gondolo:1990dk}, where $x=m/T$ and $K_i$ are the modified Bessel functions of order $i$, and
$\langle \sigma v\rangle $ can be written in terms of a total integral
\begin{eqnarray}
\langle \sigma v \rangle = {1\over 8 m_a^2 T K_2^2 (m_a/T)} \int_{4m_a^2}^{\infty} \sigma(\hat s -4 m_a^2  ) \sqrt{s} K_1(\sqrt{s}/T) ds
\end{eqnarray}
Since the freeze-out of the cold dark matter occurred when WIMP is non relativistic, one can approximate $\langle \sigma v\rangle $ with the non-relativistic expansion $\langle \sigma v \rangle  \approx a + b \langle v^2 \rangle$, where $v\equiv v_{\rm lab}$. 
For our model the thermal average of the reduced cross section can be written as
\begin{eqnarray}
\langle \sigma v \rangle (aa\to ss) &= &{\lambda_s^2 \sqrt{ m_a^2 -m_s^2 } \over 16 \pi m_a^3 }  - { \lambda_s^2 (4m_a^2 -5m_s^2 ) \over 128 \pi m_a^3 \sqrt{m_a^2 -m_s^2 }} \langle v^2 \rangle \; ,   \\
\langle \sigma v \rangle (aa\to h h) &=& {\lambda_s^2  s_\theta^2\sqrt{ m_a^2 -m_h^2 } \over 16 \pi m_a^3 }  - { \lambda_s^2 s_\theta^2 (4m_a^2 -5m_h^2 ) \over 128 \pi m_a^3 \sqrt{m_a^2 -m_s^2 }} \langle v^2 \rangle \; , \\
\langle \sigma v \rangle ( a  a \to  h  s) &= & {\lambda_s^2 s_\theta^2 \lambda^{1/2}(4m_a^2, m_h^2, m_s^2) \over 128 \pi m_a^4}\nonumber \\&&-{\lambda_s^2 s_\theta^2 [3 \lambda(2m_a^2, m_h^2, m_s^2) +4 m_a^2 (m_a^2-m_h^2-m_s^2)] \over 1024 \pi m_a^4 \lambda^{1/2} (4m_a^2,m_h^2,m_s^2)} \langle v^2 \rangle \; , 
\end{eqnarray}
where $\lambda(x,y,z)=x^2+y^2 +z^2-2xy-2yz-2xz$, $s_\theta=\sin \theta$ and $\langle v^2 \rangle =6/x_F$, with $x_F$ being the freeze-out temperature, which can be estimated through the iterative solution of the following equation ~\cite{Bertone:2004pz}
\begin{eqnarray}
x_F=\ln \left[  c(c+2) \sqrt{45\over 8} {g_*\over 2\pi^3}{ m_{DM} M_{pl} (a+6b/x_F) \over \sqrt{g_* x_F}}\right]
\end{eqnarray}
where $c$ is an order one constant. 

The final relic density can be expressed in terms of the critical density~\cite{Bertone:2004pz}
\begin{eqnarray}
\Omega h^2 \approx {1.07 \times 10^9 {\rm GeV^{-1}} \over M_{pl}} {x_F \over \sqrt{g_*} } {1 \over a+3 b/x_F} \label{relicexpress}
\end{eqnarray}
where $M_{pl} =1.22\times 10^{19}$, being the Planck mass; $g_*$ is the degree of the freedom at the freeze-out temperature.

\begin{figure}
  \includegraphics[width=0.45\textwidth]{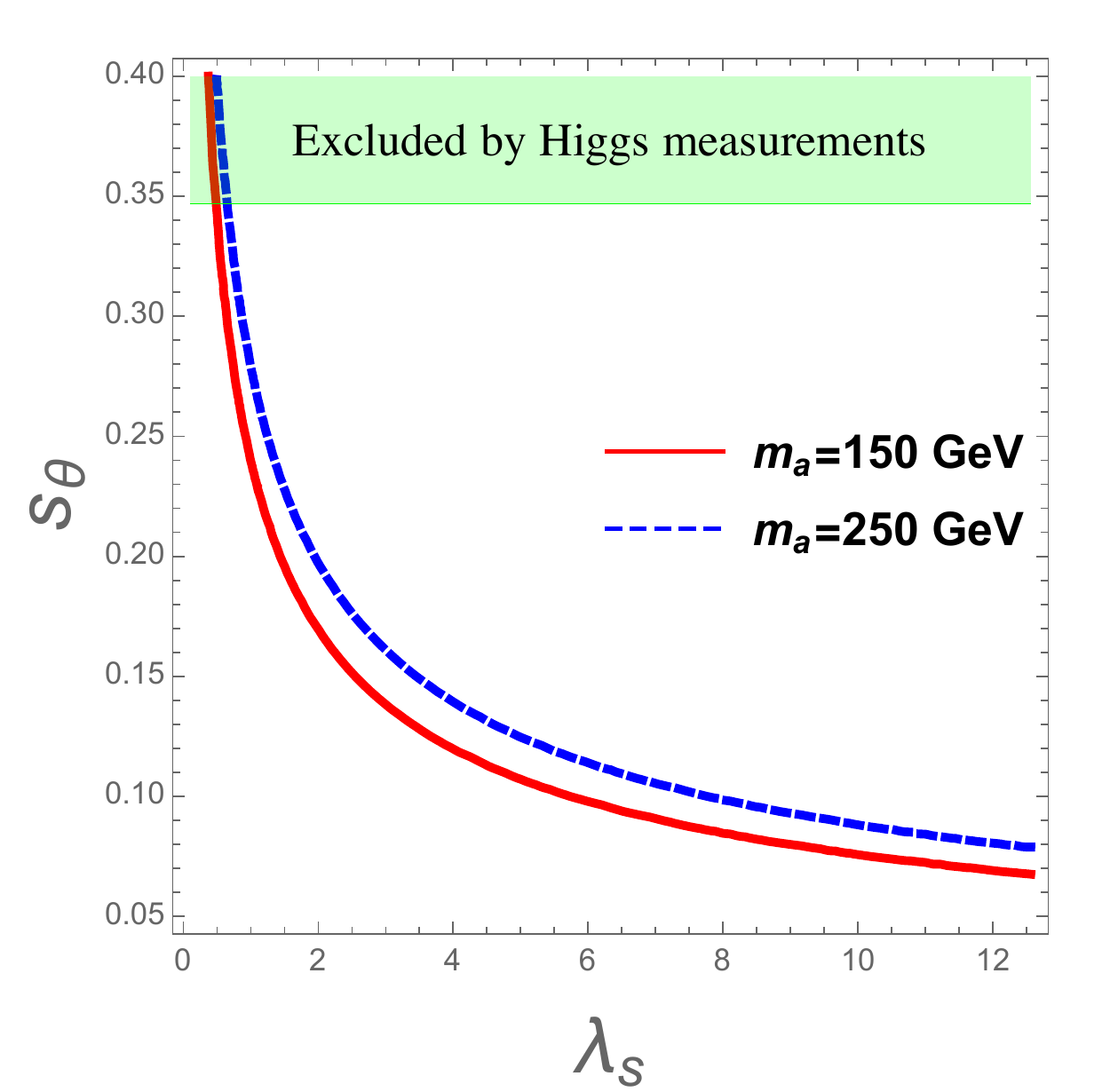}
  \includegraphics[width=0.45\textwidth]{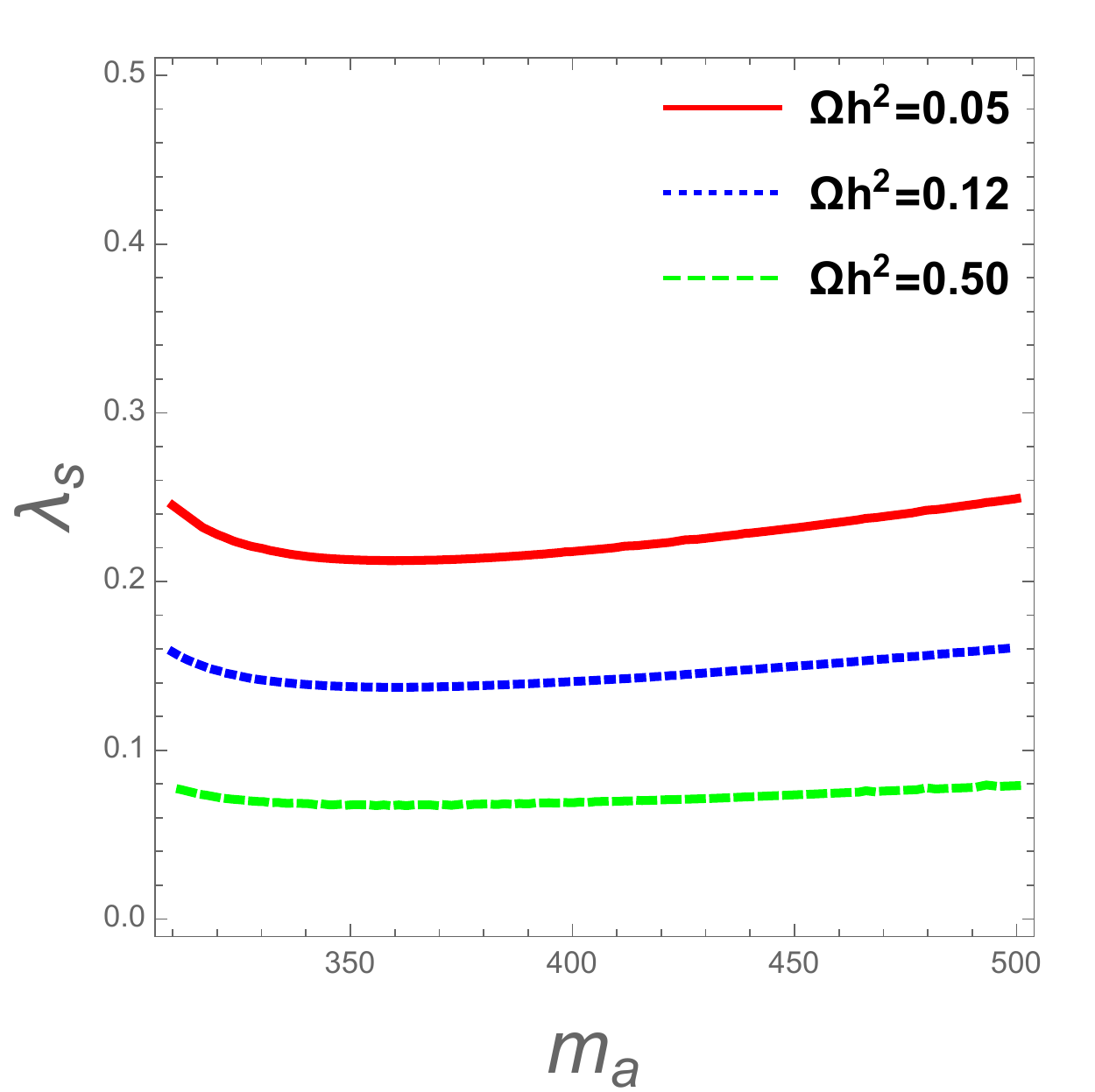}
\caption{\label{relic1} Contours of the dark matter relic density in the $s_\theta-\lambda_s $ plane by setting $m_s=300$ GeV.
}
\end{figure}

Given the Eq. (\ref{relicexpress}), one can carry out numerical analysis.
We first assume $a$ and $s$ are heavier than the SM Higgs.
For the scenario $m_h<m_a < (m_s +m_h)/2$, dark matter mainly annihilates into $2h$, which decay subsequently. 
In Fig.  \ref{relic1}, we show contours ($\Omega h^2 =0.1189$) of the dark matter relic density in the $s_\theta - \lambda_s $ plane by setting $m_s=300~{\rm GeV}$. 
The solid and dashed lines correspond to $m_a=125~{\rm GeV}$ and $250~{\rm GeV}$ respectively.  
The green band is excluded by the Higgs measurements.  It is clear that $\lambda_s$ should be significant to get a correct relic density in this case.
For the scenario $(m_s+m_h)/2<m_a <m_s$, dark matter dominantly annihilate into $\hat h\hat s$ and the dark matter relic density is less sensitive to the mixing angle $\theta$, as can be seen from the left panel of Fig. \ref{relic1}.
For the scenario $m_a > m_s$, the dark matter mainly annihilates into $ss$ final state and the relic density is not sensitive to $s_\theta$. 
We show in the right panel of Fig. \ref{relic1} contours of the WIMP relic density in the $\lambda_s-m_a$ plane. 
The scalar  self coupling $\lambda_s$ can be very small to get a correct relic density in this scenario.

\begin{figure}
{
   \includegraphics[width=0.15\textwidth]{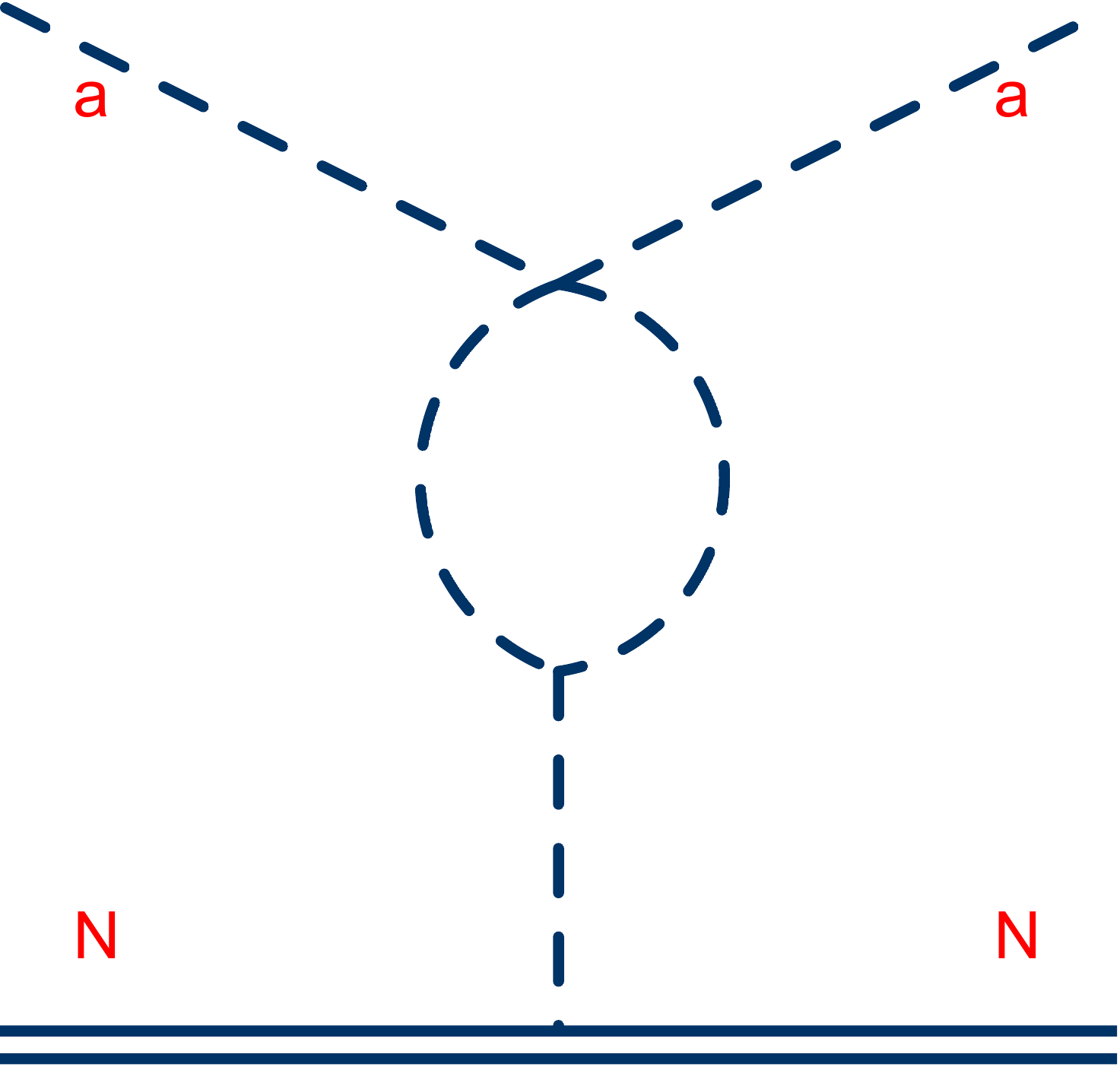}
   \hspace{2cm}
  \includegraphics[width=0.15\textwidth]{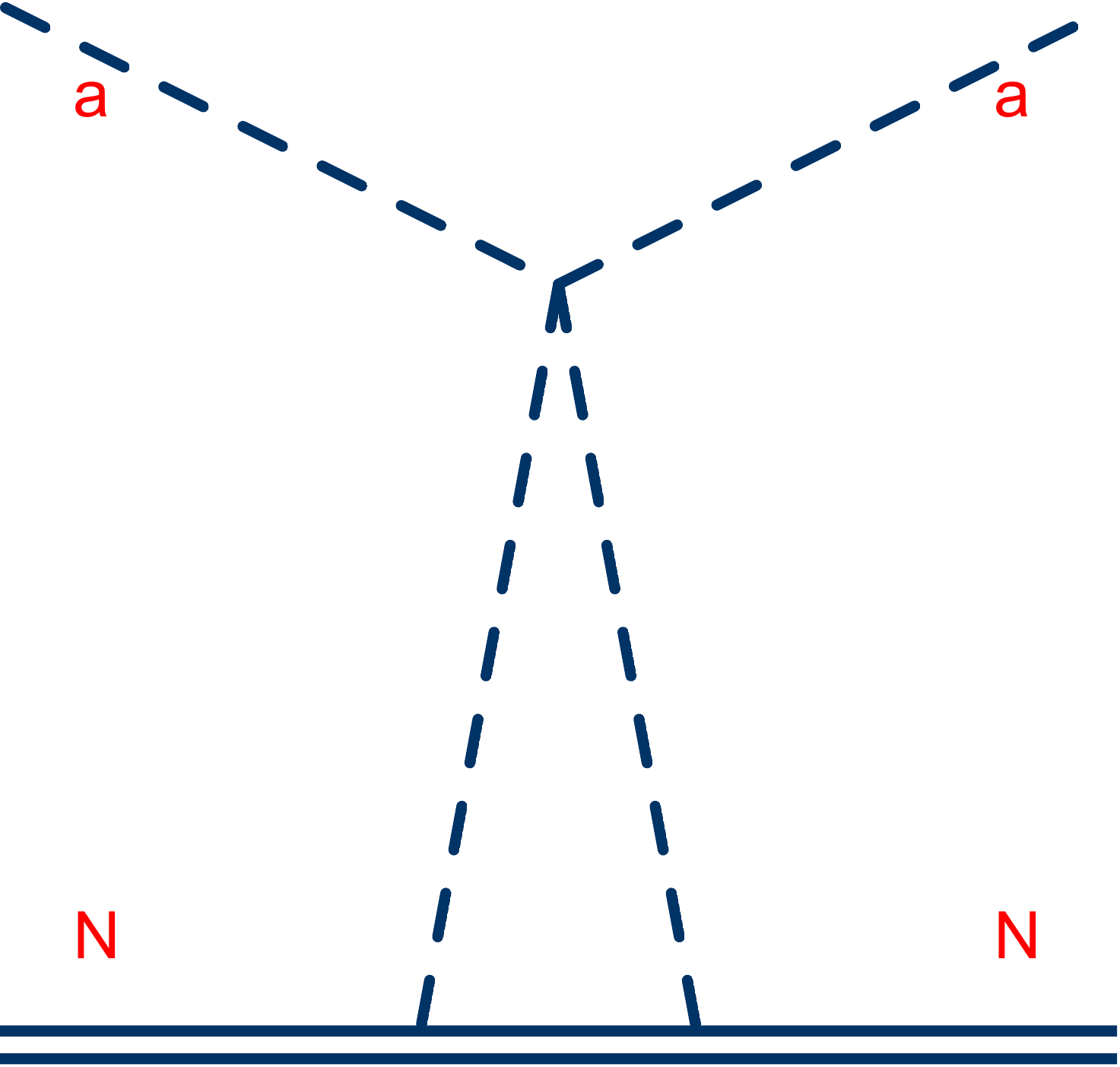}
   \\
   (a) \hspace{4cm}(b)
}
\caption{\label{ddfeyn} Feynman diagrams relevant to the direct detection of the dark matter.
}
\end{figure}

For conventional Higgs portal dark matter models,  constraints from dark matter direct detection are very severe. 
Most of parameter space of these models is excluded by the  limit  given by the LUX experiment.
For our model,  the  $t$-channel  Higgs mediated diagram, for the scattering of $a$ off nucleus, is negligible. 
The scattering happens only at the one-loop level. 
Relevant Feynman diagrams are given in Fig. \ref{ddfeyn}, (a) and (b). 
The contribution of Fig. \ref{ddfeyn} (b) is double suppressed by quark Yukawa couplings, such that we only consider the effect of  Fig. \ref{ddfeyn}, (a).
The effective Hamiltonian of the WIMP-quark interaction can be written as
\begin{eqnarray}
{\cal L} \sim { \zeta\over  m_h^2} a^2 \bar q m_q^{}  q \; ,
\end{eqnarray}
with
\begin{eqnarray}
\zeta &=& {\lambda_s\over 16 \pi^2 } \left\{c^2 \left[ 6 \lambda_{h} cs^2 +\sqrt{2} \alpha (2cs^2-s^3){ 1\over v}\right] \ln{m_s^2 \over \Lambda^2 }   + s^2 \left( 6c^3 \lambda_{h} -3\sqrt{2} sc^2 {\alpha \over v}  \right) \ln{m_h^2 \over \Lambda^2 } \right. \nonumber \\
&&-\left. cs \left[ 6c^2 s \lambda_{h} +\sqrt{2} (c^3-2cs^2) {\alpha \over v}\right] \left(-1+\ln{m_h^2 \over \Lambda^2} +{m_s^2 \over m_s^2 -m_h^2} \ln  {m_s^2 \over m_h^2}\right) \right\}
\end{eqnarray}
where $\Lambda$ is the cut-off scale, which is assumed to be $1~{\rm TeV}$ in our analysis. 
The nucleonic matrix element is parameterized as  $\langle N | \sum_q m_q \bar q q |N\rangle =m_N f_N$~\cite{Jungman:1995df}, where $m_N$ is the nuclei mass and 
\begin{eqnarray}
f^{p,n} = \sum_{i=u,d,s}  f_{q}^{p,n} + {2 \over 9 } \left( 1- \sum_{q=u,d,s} f_q^{p,n} \right)  \; .
\end{eqnarray}
Here $f^{p,n}$  are form factors of proton and neutron, which approximately equal to $ 0.283$~\cite{Belanger:2013oya}. The total cross section is 
\begin{eqnarray}
\sigma ={\mu^2 \over  \pi } \left( \zeta\over  m_a^{} m_h^2 \right)^2  \left( Z m_p f^p + (A-Z) m_n f^n  \right)^2 
\end{eqnarray}
where  $\mu = m_a m_N/ (m_a+m_N)$ being the reduced mass of the WIMP-nucleon system and $\sigma$ is the spin-independent (SI) cross section. 
%

%
%
%
\begin{figure}[t]
  \includegraphics[width=0.6\textwidth]{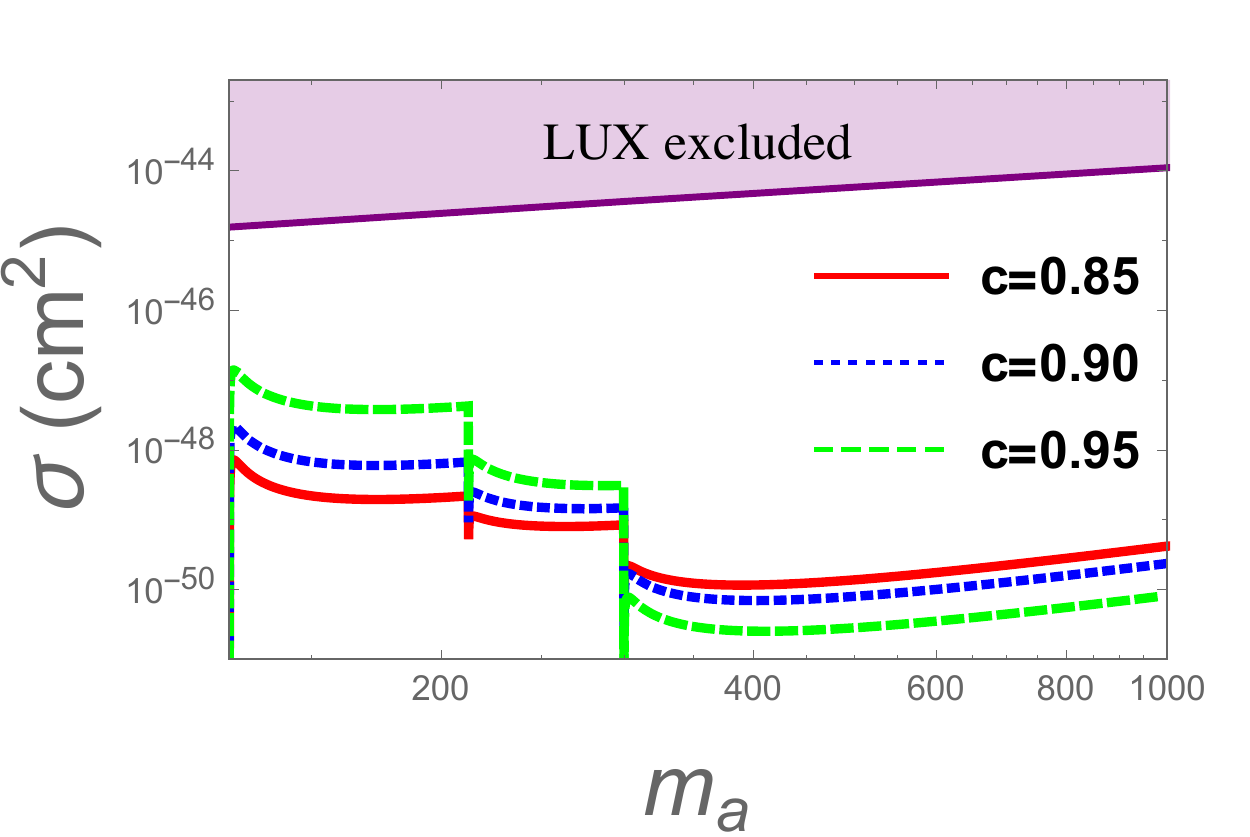}
\caption{\label{directdet} Direct detection cross section as the function of the dark matter mass.
}
\end{figure}
%
%
%

Currently the strongest constraint on the SI cross section comes from the LUX~\cite{Akerib:2013tjd}. 
We plot in Fig. \ref{directdet} the direct detection cross section as the function of the dark matter mass, where the solid, dotted and dashed lines correspond to $c=0.85,~0.90$ and $0.95$ respectively.  
For each curve, one has a correct dark matter relic density, which controls the shape of the curve.
The purple solid line is the LUX limit and the region above the line is excluded. 
Notice that our prediction of SI cross section is about $2 \sim 5$ orders below the current LUX limit, which means less possibility of detecting this types of WIMP in the underground laboratory by comparing with other conventional dark matter models, especially when the neutrino background is taken into account.
One needs to pursue the signature of the model at colliders.

\section{Collider searches}

\begin{figure}[t]
 \includegraphics[width=0.45\textwidth]{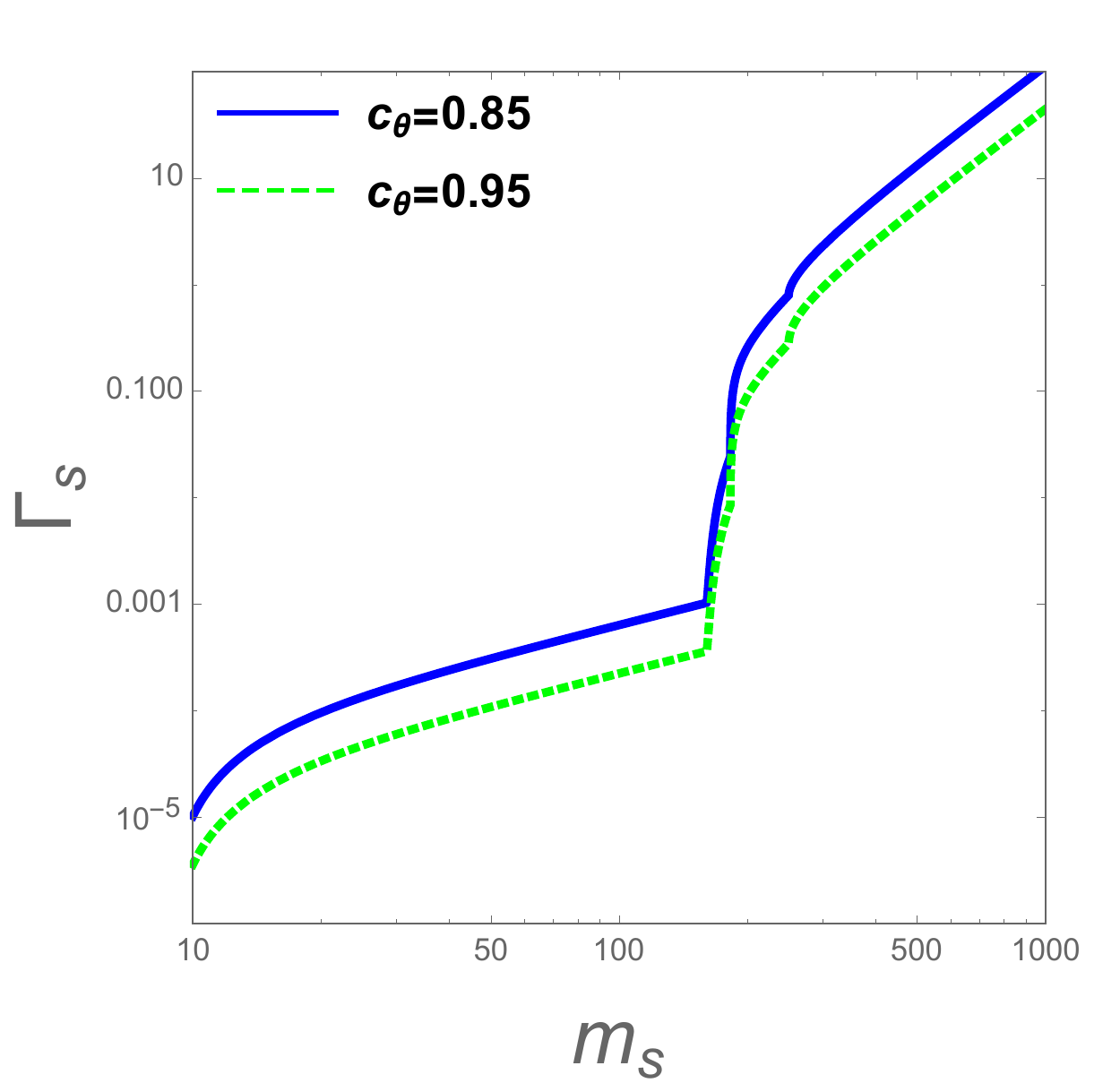}
  \includegraphics[width=0.45\textwidth]{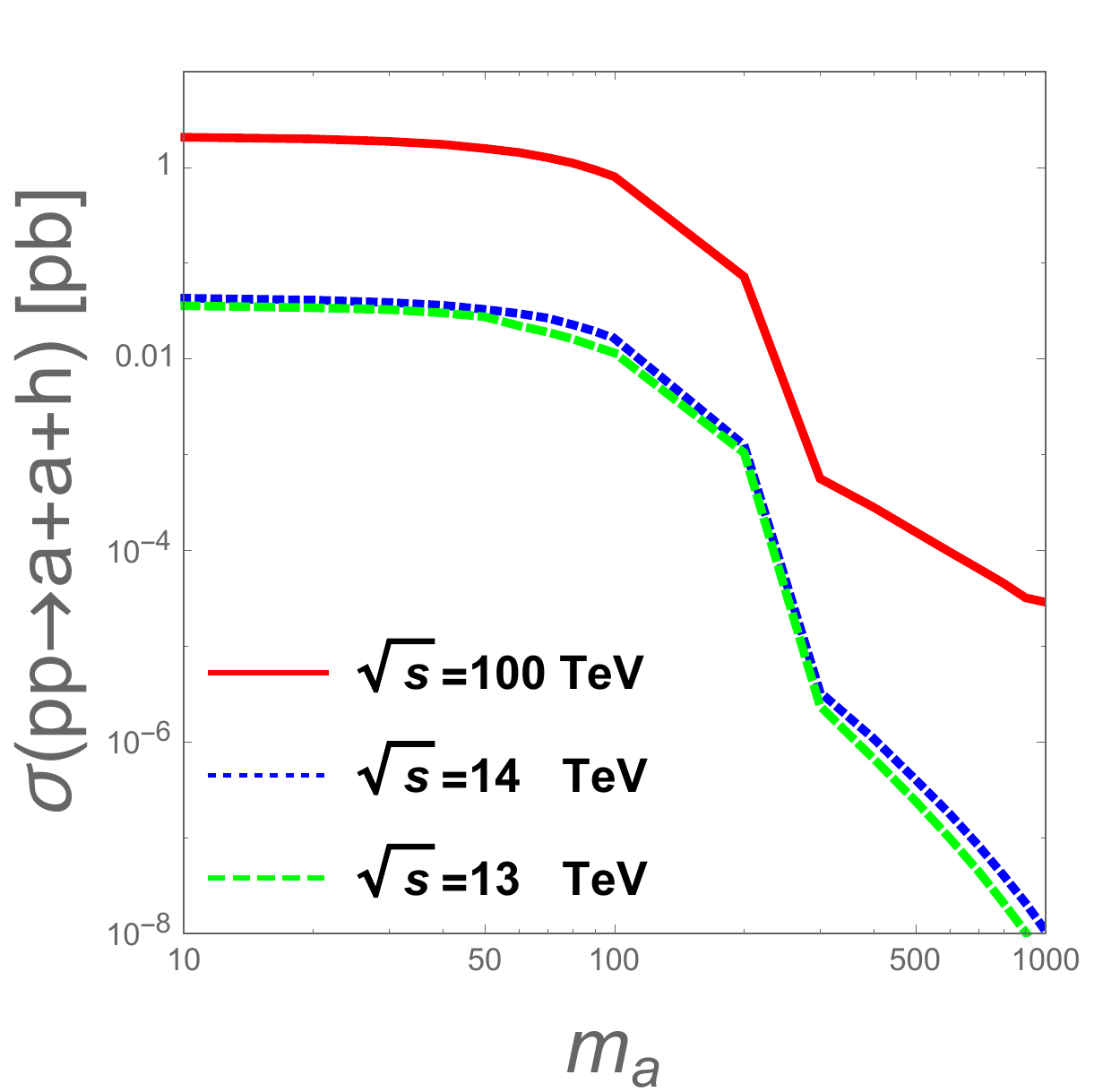}
\caption{\label{collider} Production cross section of $\sigma(pp\to a+a+ h)$ at colliders.
}
\end{figure}

At the LHC, the signature of Higgs portal dark matter  models comes from mono-Higgs searches~\cite{Carpenter:2013xra,Berlin:2014cfa}, where the dark matter are pair produced in association with a SM Higgs boson and the signature is a single Higgs boson plus missing energy from the dark matter. 
It was clarified in Ref.~\cite{Carpenter:2013xra} that the diphoton final state, which comes from the decay of the SM Higgs, provides the best sensitivity when performing an LHC background study. 
%
%
Since there will be  resonance  enhancements to the production cross section whenever the mediators are produced on-shell, it makes sense to know the decay behavior of the SM-Higgs $h$ and scalar singlet $s$.  
The decay rate of the SM Higgs is simply rescaled by a factor of $c_\theta^2$, i.e., $\Gamma_{h} \sim c_\theta^2 \times 0.0041$ assuming  a negligible $\Gamma(h\to 2s)$. 
We show in the left panel of Fig.~\ref{collider} the decay rate of $s$ as the function of $m_s$ with the solid and dotted lines correspond to $c_\theta =0.85$ and $0.95$ respectively.

Signal events can be generated using MADGRAPH~\cite{Alwall:2014hca} with showering and hadronization by PYTHIA~\cite{Sjostrand:2007gs} and detector simulation by DELPHES~\cite{deFavereau:2013fsa} at the pp collider. 
In the right panel of Fig.~\ref{collider} we show the production cross sections of $\sigma(pp\to a+a+h)$ at the LHC, where the solid, dotted and dashed lines correspond to $\sqrt{s} =100~{\rm TeV}, 14~{\rm TeV}$ and $13$ TeV respectively.
We have assumed that $m_s=500~{\rm GeV}$, $\lambda_s =2$ and $c_\theta=0.85$ when making this plot.
The best signal of this model is $\gamma \gamma +\slashed{E}_T$, with $\slashed{E}_T$ the missing energy. For the background and event selection, we refer the reader to Ref.~\cite{Carpenter:2013xra} for detail.
 %
%
A systematic analysis of collider signatures of this model will be given in a future study. 

%
%
%
%
%
%
%
%
%
%

\section{Conclusion}
We investigated a Higgs portal scalar dark matter scenario, where the dark matter is the CP odd component of a complex scalar singlet  $S$ that gets no VEV but its CP-even component mixes with the SM Higgs boson. 
The  potential has a partially broken $Z_2$ symmetry:  interactions of the CP-odd scalar keeps the $Z_2$ symmetry  while interactions of CP-even scalars  explicitly break this symmetry, which is the origination of the mass splitting between the CP-even scalar and the dark matter.  
The model contains five free parameters: $m_s$, $m_a$, $m_h$, $\theta$ and $\lambda_s$, other parameters are either negligible or reconstructable by these free parameters . 

We studied  constraint on the parameter space of the model from Higgs measurements, which has $|\theta|<0.526$ by performing a universal Higgs fit to the data of ATLAS and CMS. 
Constraints from Higgs to invisible decays and hints from heavy Higgs decays were also presented.
For the dark matter sector,  only the quartic scalar coupling is relevant to the relic density. 
One free parameter can be eliminated by requiring  a correct relic abundance. 
Besides, there is no tree level contribution to the WIMP-nucleus scattering cross section, which greatly loose the tension of the correct relic density with constraints of dark matter direct detections.   
This is the key point of our model. 
Numerical results show that the direct detection cross section is about $2\sim 5$ orders below the current LUX limit.  
Although this WIMP is hard to be detected at the underground laboratory,  it is still possible to search this kind of model at colliders in the mono-Higgs channel. 
We calculated the production cross section at the LHC, which has $\sigma(pp\to 2a+h)\sim 10~{\rm fb}$ for $m_a\sim100~{\rm GeV}$. 

\begin{acknowledgments}
The author is in debt to Huaike Guo and Haolin Li for their help in collider signature simulations.  The author also thanks Jonathan Kazaczuk, Michael J. Ramsey-Musolf, Peter Winslow and Jiang-hao Yu  for helpful discussion. This work was supported in part by DOE Grant DE-SC0011095. 
\end{acknowledgments}

\appendix

\section{An alternative dark matter model}
This model extends the SM with a complex singlet $\Phi$ and a real scalar singlet $S$. The Higgs potential can be written as
\begin{eqnarray}
V&=& -\mu_h^2 H^\dagger H -\mu_s^2 S^2 + \mu_\Phi^2 \Phi^\dagger \Phi + \lambda_h (H^\dagger H)^2 + \lambda_s S^4 + \lambda_\Phi (\Phi^\dagger \Phi)^2  \nonumber \\
&&+ \lambda_{sh} S^2 H^\dagger H + \lambda_{h\Phi} (\Phi^\dagger \Phi) (H^\dagger H) + \lambda_{s\Phi} (\Phi^\dagger \Phi) S^2 \nonumber \\
&&+ \left\{ \mu_{A}^2 S\Phi +\mu_B^2 \Phi^2 + \lambda_C  S^3 \Phi  + \lambda_D S^2 \Phi^2 + {\rm h.c.}\right\}
\end{eqnarray}
where all couplings are real and $\Phi=(\rho + v_\Phi + i A )/\sqrt{2}$ and $S=s+v_s$.  Notice that $V$ is incomplete and we leave the analysis of the complete potential for a future study. The tadpole conditions can be written as
\begin{eqnarray}
&&-\mu_h^2  + \lambda_h v_h^2 +  \lambda_{sh} v_s^2   +{1\over 2 } \lambda_{h\Phi} v_\Phi^2 =0 \\
&& (-2\mu_s^2 +4\lambda_s v_s^2 + \lambda_{sh} v_h^2 + \lambda_{s\Phi} v_\Phi^2 )v_s+ \sqrt{2} \mu_A^2 v_\Phi+ 3\sqrt{2} \lambda_C v_\Phi v_s^2 + 2\lambda_D v_s v_\Phi^2 =0 \\
&&(\mu_\Phi^2 +2 \mu_B^2 + \lambda_\Phi v_\Phi^2 + {1\over 2 } \lambda_{h\Phi} v_h^2 + \lambda_{s\Phi} v_s^2 +2\lambda_D v_s^2 )v_\Phi  + \sqrt{2} \mu_A^2 v_s  + \sqrt{2} \lambda_C v_s^3 =0 
\end{eqnarray}
The condition for the $v_\Phi =0 $ is 
\begin{eqnarray}
\mu_A^2 + \lambda_C v_s^2 =0 \; .
\end{eqnarray}
The vacuum expectation values can be written as
\begin{eqnarray}
v_h^2 = {4 \lambda_s \mu_h^2 - 2 \lambda_{sh} \mu_s^2 \over 4 \lambda_s \lambda_h - \lambda_{sh}^2 } \; ,  \hspace{0.5cm}
v_s^2 = {2 \lambda_h \mu_s^2 - \lambda_{sh} \mu_h^2 \over 4 \lambda_s \lambda_h - \lambda_{sh}^2 } \; .
\end{eqnarray}
The scalar mass matrix turns out to be 
\begin{eqnarray}
{\cal M}^2 =\left( \matrix{2 \lambda_h v_h^2 & 2 \lambda_{sh } v_s v_h & 0 \cr \bigstar & 8\lambda_s v_s^2 & 2\sqrt{2 } \lambda_C v_s^2 \cr \bigstar & \bigstar & \mu_\Phi^2 +2 \mu_B^2 + {1\over 2} \lambda_{h\Phi} v_h^2 + \lambda_{s\Phi} v_s^2 +2 \lambda_D v_s^2   }\right)
\end{eqnarray}
which can be diagonalized by the $3\times 3$ unitary transformation.
The mass eigenvalue of the CP-odd scalar is 
\begin{eqnarray}
m_A^2=\mu_\Phi^2 -2 \mu_B^2 -s\lambda_D v_s^2 + \lambda_{s\Phi} v_s^2 + {1\over 2 } \lambda_{h\Phi} v_h^2 \; . 
\end{eqnarray}
It is a dark matter candidate in this model.

\end{document}